\documentclass[
12pt,
3p,
review,
]{elsarticle}
\usepackage{lineno,hyperref}
\modulolinenumbers[1]
\usepackage{graphicx,graphics}
\usepackage{placeins}
\usepackage{amsmath}
\usepackage{amssymb}
\usepackage{mathtools}
\usepackage{array}
\usepackage{lineno,hyperref}
\usepackage{color}
\usepackage{soul}
\usepackage{dcolumn}% Align table columns on decimal point
\usepackage{bm}% bold math
\usepackage{todonotes}
\usepackage[utf8]{inputenc}
\usepackage[english]{babel}
\usepackage{libertine}
\usepackage{pgfplots}
\usepgfplotslibrary{polar}
\pgfplotsset{compat=1.9}
\usetikzlibrary{calc}

\usepackage[title]{appendix}

\usepackage{multirow}

\usepackage{layouts}

\usepackage{caption}
\usepackage{subcaption}
\modulolinenumbers[5]

\renewcommand{\l}{\mathopen{}\mathclose\bgroup\left}
\renewcommand{\r}{\aftergroup\egroup\right}

\let\originaleps=\epsilon
\let\epsilon=\varepsilon
\let\varepsilon=\originaleps
\usepackage{stmaryrd}

\mathchardef\hy="2D
\definecolor{mydark_blue}{RGB}{0, 0, 139}
\definecolor{myblue}{RGB}{0, 0, 255}
\definecolor{mycyan}{RGB}{0, 255, 255}  
\definecolor{mygreen}{RGB}{0, 255, 0}
\definecolor{myyellow}{RGB}{255, 255, 0}
\definecolor{myred}{RGB}{255, 0, 0}
\definecolor{mydark_red}{RGB}{139, 0, 0}
\definecolor{myblack}{RGB}{0, 0, 0}

\definecolor{BRY_1}{RGB}{  0,  0,255}
\definecolor{BRY_2}{RGB}{127,  0,127}
\definecolor{BRY_3}{RGB}{255,  0,  0}
\definecolor{BRY_4}{RGB}{255,127,  0}
\definecolor{BRY_5}{RGB}{255,255, 85}

\journal{XXXXXX}

\begin{document}

\begin{frontmatter}
\title{Befriending ChatGPT and other superchatbots: An AI-integrated take-home assessment preserving integrity}

% \author[mymainaddress]{Mridhula Venkatanarayanan}
\author[mymainaddress,mysecondaryaddress]{P G Kubendran Amos\corref{mycorrespondingauthor}}
\ead{prince@nitt.edu}

\cortext[mycorrespondingauthor]{P G Kubendran Amos}

\address[mymainaddress]{Theoretical Metallurgy Group,
Department of Metallurgical and Materials Engineering,\\
National Institute of Technology Tiruchirappalli, \\
Tamil Nadu, India}

\address[mysecondaryaddress]{Institute of Applied Materials (IAM-MMS),
Karlsruhe Institute of Technology (KIT),\\
Strasse am Forum 7, 76131 Karlsruhe, Germany
}

\begin{abstract} 

With the launch of ChatGPT, serious concerns have reasonably been raised of its ill-effect on the integrity of  remote take-home exams. By way of mitigating the concern, in this study, a rather straightforward Artificial-Intelligence (AI)-integrated take-home assessment technique is proposed, and the outcome of its practice is discussed. Despite involving AI, in the form of ChatGPT, the assessment adheres to the convention of posing questions invoking critical thinking and problem solving skills. However, AI is characteristically integrated in this assessment by instructing the learners to employ ChatGPT as one of the primary sources. The learners are directed to report the use of ChatGPT by including both the prompts and its responses, before expressing their thoughts on AI-generated answers and their own concluding statement. These three characteristic components of the present techniques - the handling of ChatGPT through the prompts, comments on the AI-responses and the concluding thoughts -  are evaluated to gauge the learning. 

The proposed assessment was assigned as a take-home group activity for a batch of seventy eight students, divided into thirteen groups. Despite addressing the same questions, there was no significant overlap in the answers. Moreover, a wide range of approaches were adopted by the groups in handling ChatGPT, which in-turn rendered different responses, ultimately drawing distinct answers. Besides preventing the undesired use of ChatGPT by explicitly integrating it, the proposed assessment seemingly helped the learners question the accuracy of its responses. This self-realised skepticism can be expected to curtail blatant malpractices involving ChatGPT in the long run.   

\end{abstract}

\begin{keyword}
ChatGPT, Take-Home Exams, Evaluation, Assessment, Integrity, Malpractice
\end{keyword}

\end{frontmatter}

% \linenumbers

\section{Introduction}

Communicating with one another is an act we humans take for granted. However, this feat became rather effortless over millennia. The ability to communicate comfortably and, for the most part, unambiguously is deemed as a milestone in human evolution~\cite{christiansen2003language}. Parallels can be drawn between the evolution of human communication and our interaction with machines, particularly computing machines. Over decades,  for many, computers have been a sophisticated machine with apparently huge potential, which can only be exploited if communicated appropriately. Stated otherwise, this selfless and generous companion had a language problem.
% Consequently, in order to exploit the complete potential of a computer, it was rather the responsibility of the user to learn and communicate properly. All this amidst the lack of effort from the other side.  
Therefore, it was wholly the responsibility of the user to communicate properly with almost no attempt made by the computer to make itself accessible~\cite{wexelblat2014history} . 
Experts skilled in communicating with the computers, in appropriate \textit{language}, made their features available to us by creating comprehensible \textit{interfaces}~\cite{barnes2010user}. These interfaces acted as our window of communication with the computers. Though this advancement allowed significantly many to access the computer successfully, it was largely made possible through human efforts with the machines seemingly reluctant to pick-up on our \textit{natural languages}. Taking cognisance of this one-way struggle, and the immense opportunity that lay ahead, if the computers understood and responded to our natural languages, experts began the attempts of incorporating \textit{intelligence} in the machines~\cite{benko2009history}. These attempts ultimately resulted in what we today refer to as the \textit{artificial intelligence}~\cite{haenlein2019brief}.

After few decades of its conception, the artificial intelligence is presently embodied in numerous fashions~\cite{brunette2009review}. Despite these various manifestations, and true to its history, the perception of artificial intelligence is undeniably realised when a machine, after comprehending a statement in our natural language, offers a response that is both relevant and semantically coherent. One such expression of intelligence is demonstrably evident in a recent advancement called \textit{ChatGPT}~\cite{chatgpt}. Structured, rather plainly, as a chatbot, the ChatGPT is considerably more. Early chatbots were largely part of interfaces serving a particular purpose of making its content more accessible~\cite{adamopoulou2020chatbots}. Indicative of its time, these chatbots often rendered null response or concession statement, if the queries were not structured properly~\cite{shawar2007chatbots}. Powered by a large language model adhering to the Transformer architecture with 175 billion parameters, and trained on colossal data, ChatGPT stands exceedingly tall as an almost perfect expression of artificial intelligence. Reflecting its complex underpinnings, the response offered by the ChatGPT to a given question ( or prompt) is generally indistinguishable from a human reaction. This convincing performance of ChatGPT has even left many experts bemused~\cite{roose2022brilliance}.

The continued realisation of the genius of ChatGPT, and its plethora of applications, is increasingly accompanied by the blaring noise of the alarm bells. While thoroughly accepting its extended utility, the alarming reports warn us of the undesired ill-effects of our advanced AI companion. Though the concern over the plausible misuse of the ChatGPT stem from several sectors, including press, legal, public relations, medicine and defence~\cite{taecharungroj2023can, li2023dark, gravel2023learning, renaud2023chatgpt, kitamura2023chatgpt}, the noise from one particular community has been deafening~\cite{ kasneci2023chatgpt,halaweh2023chatgpt}. This community, primarily comprising of educators across different levels and varied streams, envisions an objectionable involvement of ChatGPT in education, and its assessments, in particular~\cite{rudolph2023chatgpt}.

% These reports stem from wide range of domains including press, legal, medicine and defence. In the midst of the concerns raised over the plausible misuse of the ChatGPT in various spheres, noise from a particular community continues to remain rather deafening. This community primarily comprises of educators across different levels and varied streams with their concerns largely pertaining to assessments.

The importance apportioned to teaching, in an educational system, is equally extended to the assessment or the evaluation of the learner's grasp~\cite{brown2013assessing}. The assessment, as the educational measurement, serves multiple purposes beginning with gauging the understanding of the learners on a specific concept. Moreover, the employers use the grades resulting from the evaluation as one of the criterion for identifying suitable candidate~\cite{hernandez2009graduates}. As a double-edged sword, a well-structured assessment would also facilitate in ascertaining the success of a teaching technique~\cite{darling2000authentic}. A huge range of schemes have been devised and adopted over different disciplined to assess the learners~\cite{grossman2014test}. Of these different assessment techniques, in addition to in-class evaluation, take-home exams are broadly practised. With its relaxed time limit, access to unlimited information, and the absence of invigilators, the take-home exams absolutely contrast the in-class exams. Despite its characteristic features apparently rendering the term exam meaningless, the involvement of take-home evaluation is convincingly justified. An in-class exam, owing to its inherent restrictions, primarily tests the information-retrieval capability of the learner. Given these proctored assessments ensure ethical practices, they also largely overlook other aspects of learning including understanding, applying, analysing, evaluating and creating, which constitute the higher levels in \textit{Bloom’s hierarchical taxonomy}~\cite{krathwohl2002revision}. On the other hand, through appropriate questions, in take-home exams, a learner is guided to define a problem, analyse it, hypothesise, validate and conclude. Furthermore, the extended time limits in take-home assignments makes room to accommodate almost the entire syllabus~\cite{bengtsson2019take}.

The foundation of the take-home exams, that involves combing through the vast pieces of information, scattered over the internet and books, to formulate a single coherent response to a question, is substantially weakened by a \textit{superchatbot}, like ChatGPT, capable of covering all these key facets, and more, in a matter of seconds. This palpable threat to a decades-old assessment technique is central to the alarm raised by the educators against ChatGPT and other similar superchatbots. In this work, besides compiling and categorising possible solutions, an AI-integrated assessment is proposed as way of safeguarding the foundation and integrity of the take-home exams. Moreover, the fascinating outcome that emerged from the trial run of this evaluation technique is discussed.

\section{Overview of plausible solutions}

\subsection{Threat is real}

In spite of the complex under-structure, its appearance and a visceral realisation of it being solely a reactionary tool, might delude us into thinking the potency of ChatGPT can be bested by questions demanding critical thinking and problem solving. Better yet, instructing the learners to frame their own problem statement and offer a coherent answer appears to render the superchatbots unnecessary. However, a recent study posing critical-thinking questions, and separately prompting to generate complex problems and its corresponding answers, in topics across four disciplines (Machine Learning, Marketing, History and Education), unravelled that the response of ChatGPT is impressive and quite indistinguishable from an efficient learner~\cite{susnjak2022chatgpt}. Furthermore, this work reports ChatGPT’s responses being relevant, clear, accurate, precise, in-depth, logical  persuasive and original. Besides attesting the potency of ChatGPT, this study emphasises the urgency for exploring solutions to avoid or detect the undesired intrusion of superchatbots in take-home exams.

\subsection{Self-redeeming options}

From their preliminary considerations, experts rendered their recommendations to detect the unwelcomed involvement of ChatGPT. These included
\begin{enumerate}
 \item \textbf{Factual inconsistencies}: The intelligence of ChatGPT is primarily a reflection of its training over vast, largely online, data. This training, though ensures semantically coherent responses, the factual claims are not always reliable~\cite{zhao2023can}. Therefore, factual inconsistencies across an answer or a response might indicate the extensive involvement of ChatGPT.
 \item \textbf{Language analysis and plagiarism tools}: Sophisticated AI-based language analysis tools can be developed and employed to realise the pattern in responses generated by ChatGPT and other superchatbots. Such tools would quantitatively detect the extent of ChatGPT’s involvement in a content generating task. Even though such language analysing tool is yet to be developed, GPT-2 Output Detector is claimed to adequately detect the content emerging from ChatGPT prompts~\cite{gao2023comparing}. Moreover, existing plagiarism detection tools can be extended to uncover the undesired role of ChatGPT in take-home exams
 \item \textbf{Augmenting discussion}: Integrity in the take-come assessment can be secured with the inclusion of a interpersonal discussion or a presentation, besides the submission of the answers. Ease with which the  learner is able to translate the written answers to a discussion relates to the originality of the work.
\end{enumerate}

Interestingly, although not verbatim, the above suggestions to avoid or detect the extensive involvement of ChatGPT and other superchatbots was rendered by ChatGPT itself~\cite{cotton2023chatting}. When conscientiously prompted about the solutions to its ill-effects on education, these and several other seemingly plausible options are offered by ChatGPT. Correspondingly, the above recommendations are categorised as \textit{self-redeeming options} of ChatGPT.

\subsection{Solutions of gaps}

Ways to detect the interference of ChatGPT, in the previous section, are largely based on the characteristic features of its outcomes. The following techniques, on the other hand, limits the usage of ChatGPT by focusing on its general structure and minutiae of the underlying algorithm.

In its technical form, ChatGPT essentially recognises the pattern in the usage of words in natural languages, and constructs a coherent conversation by converting these words into tokens (number). When these token are expressed in a specific series, reflective of the pattern gleaned from a language, it essentially translates to a coherent sentence~\cite{vaswani2017attention}. This mechanism is rather reversed by ChatGPT to comprehend a given prompt.

Taking cognisance of the inner-workings the following approaches can be adopted to restrict their involvement.
\begin{enumerate}
 \item \textbf{Incorporating Images}: Owing to their fundamental framework, ChatGPT and similar superchatbots are essentially a reactionary aide that exclusively respond to verbal prompts. Therefore, a take-home assessment structured a around set of distinct images would render the superchatbots unusable.
 \item \textbf{Evasive prompts}: The huge corpus of information that constitute the training data for ChatGPT primarily includes the conventional use of language. Therefore, phrases a question in an unconventional semantic, as often observed in riddles and poems, introduces a misconception in the understanding of ChatGPT, thereby compromising the accuracy of its responses~\cite{lehman2023machine}. Questions can be phrased in these forms, which seemingly are comprehensible to humans, but evade the understanding of the superchatbots. The prompts with unconventional language schema is shown is generate inaccurate responses from ChatGPT. Besides unconventional use of language, it has also been demonstrated that ChatGPT exhibits a weak comprehension of spatial positions and temporal series of events~\cite{borji2023categorical}. Questions involving these features is prone to yield inaccurate solutions from ChatGPT which can be exploited while framing the take-home exam.
\end{enumerate}

Though the aforementioned techniques render the ChatGPT unusable and its response inaccurate, these are categorised as solutions of gaps, in that, having realised these limitations the subsequent advancements in superchatbots will largely be focused on addressing them. In other words, the efficacy of exploiting the inherent restriction of the models supporting the ChatGPT is swiftly declining and might soon be ineffective.

\section{AI-integrated take-home exam}

Even though the previous section is far from an exhaustive review of the techniques involving ChatGPT and take-home exams, it essentially reflects the prevalent sentiment among the educators.  Stated otherwise, the common consensus, as it stands, is that the only way ChatGPT and other superchatbots can be associated with a taken-home assessment is if their roles are avoided and/or detected to conserve integrity. This work, on the other hand, describes an approach for consciously involving ChatGPT, or other similar superchatbots, in take-home exams without compromising the integrity.

In fact, a technique of involving ChatGPT in evaluation has already been suggested by ChatGPT itself. In this assessment technique, ChatGPT can be employed to devise personalised assessment for individuals in the forms of an online interactive assistants,  games, and other dedicated evaluating tools. Similar thoughts have also been shared by the known online educator Sal Kahn, the creator of Kahn Academy~\cite{kahn}. Though ChatGPT makes the development of such chatbots, which can be deployed for evaluating learners on specific topics, the instructor is required to possess some level of fluency in computer languages. The proposed technique, in contrast, is founded on the traditional scheme of take-home exam.

\subsection{Theoretical framework}

Following the conventional approach, in the AI-integrated take-home assessment, questions focusing on the higher strata of Bloom’s taxonomy, that examines understanding, application, analysis, evaluation and creation, is posed to the learners. However, instead of just necessitating a well-thought-out coherent answers, the learners are instructed to use ChatGPT as a mandatory, yet not exclusive, source. It is advised to present the approach they adopted in utilising ChatGPT, by including all the prompts and responses, in its raw form, as a screenshot or equivalent. Finally, besides their concluding statement, the learners are required to comment on the responses of ChatGPT.  Simply put, the AI-integrated take-home exam by explicitly involving ChatGPT, as a primary source, requires the learners to gauge its responses, in addition to rendering their answer in view of this and other sources. By carefully considering the ChatGPT prompts, the comments on its responses and the finally, the concluding statement, suitable assessment can be made of the higher levels of Bloom’s taxonomy. 

Though the assessment-question can directly be posed to ChatGPT, a more convincing answer is generally attained when prompted further, particularly when the response demands critical thinking and problem solving skills. The subsequent prompts,  along with the original at time, aide the evaluator in gauging the understanding of the learner. Put differently, without sufficient \textit{understanding} of the topic and \textit{analysis} of the response offered to the foundational prompt, the learner would not be able to frame subsequent question to ChatGPT. Consequently, the prompts will reflect the understanding and analysis of the learners.  The comment on the responses of the ChatGPT, which is a characteristic aspect of the proposed assessment scheme, necessitates in-depth \textit{evaluation} of its content, besides analysis. Finally, the \textit{application} of the response offered by ChatGPT, along with other sources, lends itself to the \textit{creation} of the concluding statement or answer to the assigned question. By encompassing all the levels of Bloom's taxonomy, present take-home exam stands true to the convention.

\begin{figure}
    \centering
      \begin{tabular}{@{}c@{}}
      \includegraphics[width=1.0\textwidth]{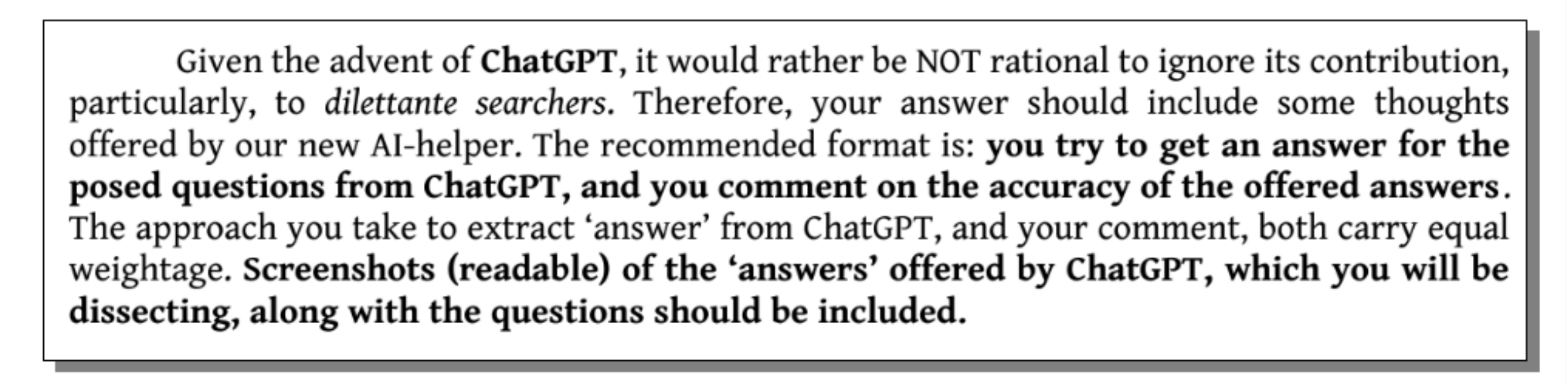}
    \end{tabular}
    \caption{  A piece of instruction included along with the questions in the AI-integrated take-home exam assigned to the learners. (This instruction was further expanded during the conventional classes to weed out any confusion.)
    \label{fig:Instruction}}
\end{figure}

\subsection{Practice}

In addition to formulating an AI-integrated take-home exam, that seemingly ensures integrity despite the access to ChatGPT, and other superchatbots, its viability is ascertained in this work by assigning it to a batch of learners. 

\subsubsection{Methodology}

The present assessment approach, characterised by the explicit involvement of ChatGPT, was the pattern adopted for the take-home assignment of the undergraduate students in their second year of the four years technical education program. These students, seventy eight in total, are pursuing bachelor's degree in the Metallurgical and Materials Engineering at National Institute of Technology Tiruchirappalli (NITT). Located in southern part of India, according to the recent official survey, the NITT ranks ninth among the technical education institutes across the nation. 

The assessment comprised of two questions pertaining to a course on Mechanical Behaviour and Testing of Materials, a core course (\textit{de rigueur}) of the discipline. 
A section of the instructions as shared with the students are shown in Fig.~\ref{fig:Instruction}. Given its uniqueness and untested nature, the assessment was structured as a group activity. Correspondingly, the students were instructed to form groups, on their own, with each comprising of six members. For the resulting thirteen groups same two questions were assigned and a deadline was set after fifteen days for the submission of the answer. (The exact questions are included in the appendix)

\begin{figure}
    \centering
      \begin{tabular}{@{}c@{}}
      \includegraphics[width=1.0\textwidth]{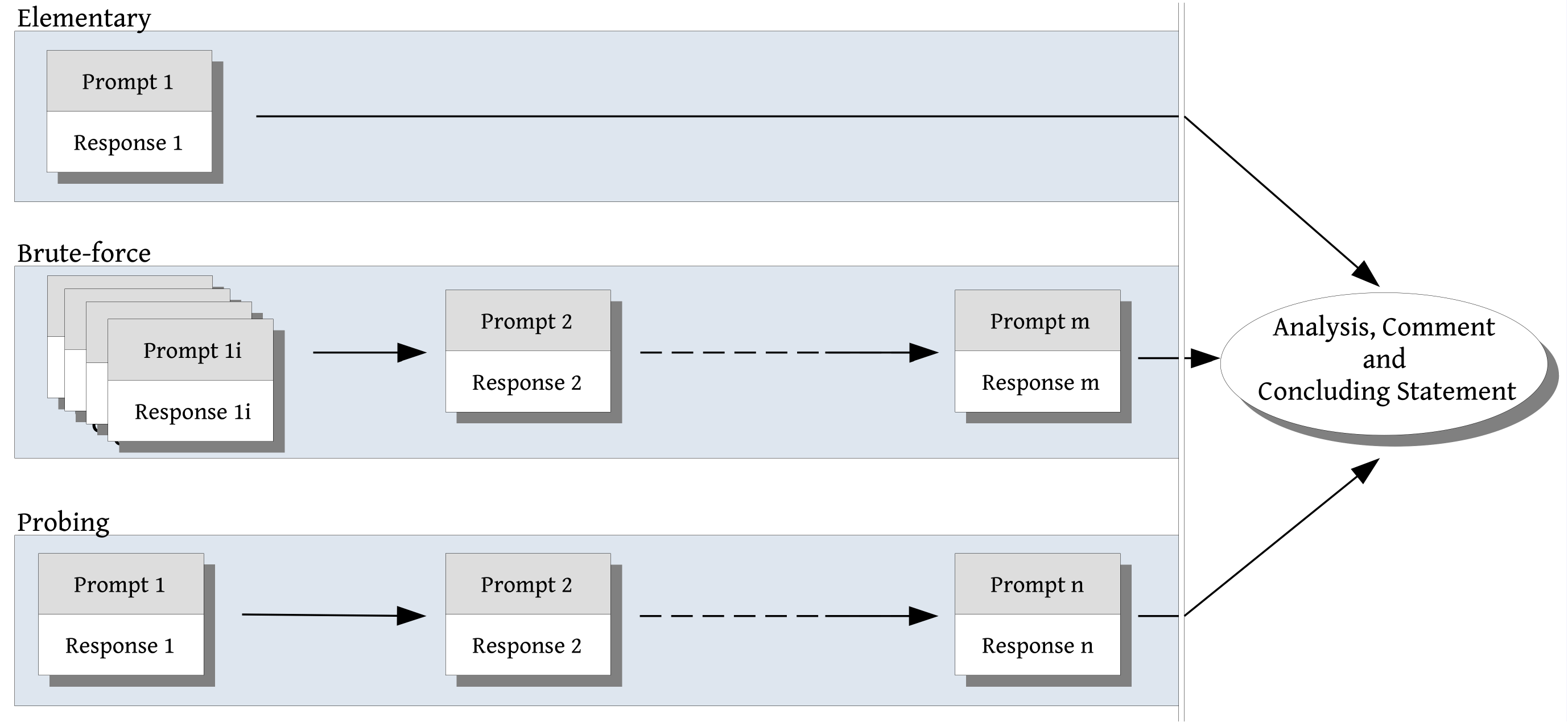}
    \end{tabular}
    \caption{  Different approaches adopted by the group in integrating ChatGPT, as instructed, to their take-home exam. Prompts 1a to 1i, in brute force approach, indicate repetitive use of the same question. Though both probing and brute force used several prompts, the number of distinct questions posed by the former is greater (n$>$m).
    \label{fig:approaches}}
\end{figure}

\subsubsection{Outcomes and discussion}

A cursory view of the responses from the thirteen groups revealed no noticeable similarities. Stated otherwise, the required involvement of ChatGPT seemingly improved the integrity of the take-home assessment, considering same questions were assigned to every group. This absence of word-to-word mirroring of answers across the groups could be attributed to the curiosity of the learners to explore the newly launched AI tool in a technical context. The varied answers are essentially indicative of the different approaches adopted by the groups is utilising ChatGPT. Though no two answers overlap completely, a similarity can be drawn in the ways ChatGPT was employed. This similarity does not suggest any malpractice, in its conventional sense. But, on the other hand, the similarities in the approaches can be exploited to succinctly categorise and describe the ways ChatGPT was handled.

The approaches adopted by the groups in involving ChatGPT can be broadly categorised into the following:
\begin{enumerate}
 \item \textbf{Elementary}: Couple of groups restricted the use of ChatGPT to once for each question. Their prompt was literally the question posed in the assignment, without any change. The ChatGPT is not prompted any further for an expansion or clarification. Despite the single use, following the instructions, the groups assessed the response of ChatGPT and made a concluding statement. A group, which did not explicitly distinguish the response from ChatGPT, is also included in this category.  Except for these, no other groups employed ChatGPT only once in answering the questions. 
 \item\textbf{Brute-force}: Adopted primarily by one group, this approach exposed ChatGPT to the same prompt over and over (close to seven times), with each generating a different response. ChatGPT was subsequently tasked to find the most appropriate answer from the aforementioned responses. Upon clarifying few aspects of this answer through suitable prompts, the group proceeded to comment on the responses of ChatGPT, and offered its own concluding statement.
 \item\textbf{Probing}:  Similar to brute-forcing, this approach extensively involved ChatGPT through numerous prompt. However, characteristically, every time different question was posed to the superchatbot. Either the prompts were re-phrased to make it more particular or to press ChatGPT for additional clarification to the answer rendered.  The prompts following the original one reflected thoughtful consideration of the response offered by ChatGPT, in view of understanding gathered from other sources. By framing the questions appropriately, ChatGPT was directed and probed, in this approach, till a convincing response was made. 
\end{enumerate}
Fig.~\ref{fig:approaches} graphically represents the different approaches adopted by the groups in integrating ChatGPT.

\begin{figure}
    \centering
      \begin{tabular}{@{}c@{}}
      \includegraphics[width=0.9\textwidth]{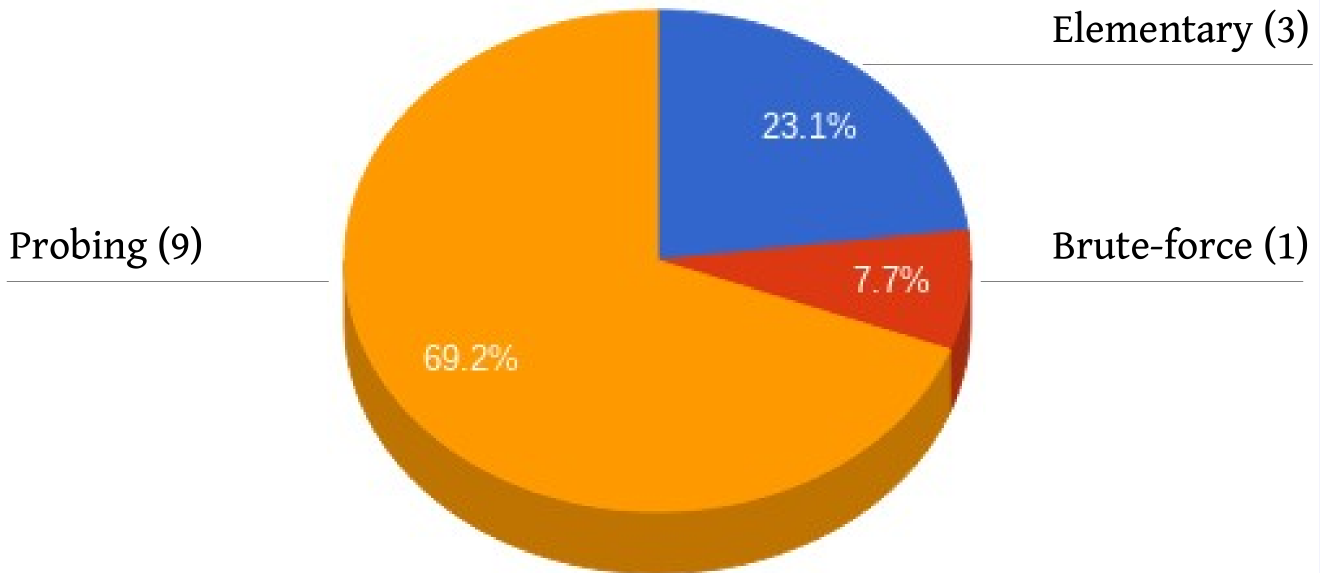}
    \end{tabular}
    \caption{  A quantative distribution of  the groups adopting varied approaches indicating major adherence probing technique.
    \label{fig:piechart}}
\end{figure}

Majority of the group adopted the probing approach to effectively use ChatGPT in their take-home assignment. Interestingly, one such group masked ChatGPT as DAN (\lq Do Anything Now\rq \thinspace) assuming it would enhance the response to their questions. The distribution of the different ways ChatGPT was employed in the present practice is graphically represented in Fig.~\ref{fig:piechart}.

As instructed, in addition to presenting their record of utilising ChatGPT, almost all groups technically assessed the content of its responses and discussed them adequately. While some groups chose to offer a consolidated evaluation of all ChatGPT's answers, each response was individually assessed by others. The concluding statement of each group largely summarised the technical assessment of the contribution of ChatGPT based on their understanding of the potential solution stemming from the lectures and/or other sources.

\section{Conclusion}

Teaching and evaluating techniques, in education, have always been evolving with time. Given the progressive changes in these techniques are induced by several factor, the corresponding evolution has not always been smooth and gradual. Sometimes, an abrupt change in the teaching and evaluating approach is rather forced. The exclusive use of online classrooms and the consequent change in the teaching and evaluation, during the recent pandemic, is a prime example of the abrupt change to the existing techniques. Though this shift to online classes was transitory, it opened-up a new avenue for distant education. The launch of ChatGPT, and other similar superchatbots, challenge some of the conventional evaluation techniques and demand rather an abrupt change for their effective continuance. The revisions made to the assessment techniques, in response to these advancements in AI, unlike pandemic, would not be transitory but, on the other hand, a landmark in the adaptive changes. In-keeping with the intelligent resources, which are accessible to the learners, an AI-integrated take-home assessment is proposed. Moreover, its practicality is demonstrated, in this work, by discussing the outcomes of the proposed evaluation technique when assigned to a batch of learners. The theory and practice of the present AI-integrated take-home exam indicates that, despite the access to ChatGPT and other superchatbots, neither the quality of the assessment nor the integrity is compromised. 

Interestingly, the concluding statement of one of the groups included:\\
\lq \dots\textit{ we can state that the  ChatGPT's accuracy is based on the applicability of the input data it (sic) has been provided. Furthermore, they might not always offer information that is accurate or trustworthy. By interpreting this, we may say that while the answers provided by the ChatGPT above are mostly right, they are not entirely accurate.}\rq \thinspace \\
wherein a generalised observation is reported, after an inquisitive treatment of ChatGPT, besides commenting on the subjective technical accuracy of the responses. Accordingly, though ChatGPT offers a seemingly coherent response, it cannot be accepted and reported, as it is, without any human assessment or intervention. Though not expected, but encouragingly, similar sentiments were expressed by several other groups in this study. These generalised claims that expresses a reluctance in accepting ChatGPT as an authoritative source is noticeable because such skepticism towards the superchatbots would implicitly avoid its involvement in any malpractices. In other words,  the present assessment scheme, in addition to ensuring integrity of take-home exams, seems to establish an understanding in learners to view the response of ChatGPT critically, and not be content with it. When the beneficiaries themselves are not convinced of a tool, its undesired usage in the midst of other established tools is rather unreasonable.

Ultimately, the progress in artificial intelligence poses a palpable challenge to education, particularly to the existing assessment techniques. However, these challenges are not insurmountable. As demonstrated in the present work, the advancements can be integrated to the existing assessment techniques, in a straightforward manner, to preclude any ill-effects, all while keeping-up the technological progress. Furthermore, such evaluations, through the first hand experience, would keep the learners well-informed of both benefits and pitfalls of the state-of-the-art technologies. 

The present study conclusively indicates that the alarms reasonably raised by the educators, envisioning the role of ChatGPT and other superchatbots, can seemingly be settled with some innovation and critical thinking, which is expected of a learner. For we, the creators, unless conceded, would always stand tall over our creations.   

\section*{Data Availability}

All the data pertinent to the reported study would be made available upon request and after consideration of the reason. 

% \section*{Acknowledgments}
% 
% PGK Amos thanks the financial support of the SCIENCE \& ENGINEERING RESEARCH BOARD (SERB) under the project SRG/2021/000092.
% % \section*{References}

\bibliographystyle{elsarticle-num}
\bibliography{library.bib}
\end{document}